\documentclass[aps,prd,twocolumn,superscriptaddress]{revtex4}
\usepackage{epsfig,epsf}
\usepackage{amsmath}
\usepackage{amsthm}
\usepackage{amsfonts}
\usepackage{amssymb}
\usepackage{dsfont}
\usepackage{multirow}
\usepackage{appendix}
\usepackage{slashed}
\usepackage[active]{srcltx}
\usepackage{psfrag}

\begin{document}

\title{Tensor molecule $J/\psi J/\psi$: A candidate to the resonance $%
X(6200) $}
\date{\today }
\author{S.~S.~Agaev}
\affiliation{Institute for Physical Problems, Baku State University, Az--1148 Baku,
Azerbaijan}
\author{K.~Azizi}
\affiliation{Department of Physics, University of Tehran, North Karegar Avenue, Tehran
14395-547, Iran}
\affiliation{Department of Physics, Faculty of Engineering and Natural Sciences, Dogus
University, Dudullu-\"{U}mraniye, 34775 Istanbul, T\"{u}rkiye}
\author{H.~Sundu}
\affiliation{Department of Physics Engineering, Istanbul Medeniyet University, 34700
Istanbul, T\"{u}rkiye}

\begin{abstract}
The hadronic tensor molecule $\mathcal{M}=J/\psi J/\psi$ is investigated in
the framework of QCD sum rule method. We evaluate its mass and current
coupling using the two-point SR approach. Our result $m=(6290 \pm 50)~
\mathrm{MeV}$ for the mass of $\mathcal{M}$ indicates that it can decay to a
pair of mesons $J/\psi J/\psi$. Apart from this dominant channel there are
subdominant modes of the molecule $\mathcal{M}$ generated due to
annihilation of constituent $\overline{c}c$ quarks to pairs of light quarks $%
\overline{q}q$ and $\overline{s}s$. This mechanism launches processes $%
\mathcal{M} \to D_{(s)}^{(\ast )+}D_{(s)}^{(\ast )-}$, $DD_{1}(2420)$, $%
D_sD_{s1}(2460)$ and $D_{(s)}^{(\ast )0}\overline{D}_{(s)}^{(\ast )0}$. The
decays of $\mathcal{M}$ are explored by applying technical tools of the
three-point sum rule approach which is necessary to estimate strong
couplings at $\mathcal{M}$-meson-meson vertices. Comparing the mass $m$ of
the molecule $\mathcal{M}$ and its decay width $\Gamma[\mathcal{M}]=(149 \pm
21)~ \mathrm{MeV}$ with available experimental data, we discuss the molecule
$\mathcal{M}$ as a possible candidate to the tensor resonance $X(6200)$.
\end{abstract}

\maketitle


\section{Introduction}

\label{sec:Intro}

The four $X$ resonances observed by LHCb-ATLAS-CMS collaborations in the di-$%
J/\psi $ and $J/\psi \psi ^{\prime }$ mass distributions with masses
covering the region $6.2-7.2~\mathrm{GeV}$ and their parameters provide
valuable information about fully charmed exotic mesons \cite%
{LHCb:2020bwg,ATLAS:2023bft,CMS:2023owd,CMS:2026tiu}. The reason is that
these structures are presumably composed of four charm quarks $cc\overline{c}%
\overline{c}$ investigation of which can shed light on properties of such
states.

Although exotic particles built of four heavy quarks were already among
priorities of theoretical studies, recent discoveries triggered new
investigations devoted to their analysis \cite%
{Anwar:2017toa,Bedolla:2019zwg,Zhang:2020xtb,Wang:2020ols,Albuquerque:2020hio,Yang:2020wkh,Cordillo:2020sgc,Dong:2020nwy,Liang:2021fzr,Wang:2021kfv,Deng:2020iqw}%
. In these publications parameters of $cc\overline{c}\overline{c}$ mesons
were calculated using all available models and methods. Such structures with
various quantum numbers were studied in the framework of the
diquark-antidiquark and hadronic molecule models which are mostly employed
ones. It is interesting that alternative interpretations of $X$ states were
proposed as well. In fact, these structures were considered in Ref.\ \cite%
{Dong:2020nwy} by utilizing a coupled-channel approach. It was demonstrated
that in the di-$J/\psi $ system exists a near-threshold state $X(6200)$
bearing quantum numbers $0^{++}$ or $2^{++}$. Due to coupled-channel effects
$X(6900)$ may also be generated as a pole structure \cite{Liang:2021fzr}.
Additionally, the authors predicted existence of a bound state $X(6200)$
with preferable quantum numbers $\ 0^{++}$, and resonances $X(6680)$ and $%
X(7200)$ with $2^{++}$, respectively.

Possible assignments of the $X$ resonances were also considered in Refs.\
\cite%
{Wang:2022xja,Faustov:2022mvs,Niu:2022vqp,Dong:2022sef,Yu:2022lak,An:2022qpt,Kuang:2023vac,Liu:2020eha,Malekhosseini:2025hyx,Song:2024ykq}%
. Thus, resonances $X(6600)$, $X(6900)$ and $X(7300)$ were analyzed as
states with quantum numbers $J^{\mathrm{PC}}=0^{++}$ or $1^{+-}$ in Ref.\
\cite{Wang:2022xja}. The author employed the diquark-antidiquark model and
used the sum rule (SR) method in conjunction with Regge trajectory
techniques. In the framework of the relativistic quark model $X(6200)$ was
explored as a ground-state scalar tetraquark $J^{\mathrm{PC}}=0^{++}$,
whereas $X(6600)$ identified with either $1S$ tensor state $J^{\mathrm{PC}%
}=2^{++}$ or a radial excitation ($2S$) of the scalar structure \cite%
{Faustov:2022mvs}. The $X(6900)$ state maybe is $2S$ excited tensor
tetraquark or $1D$ scalar/tensor exotic mesons.

The $X$ resonances were studied in our works \cite%
{Agaev:2023wua,Agaev:2023ruu,Agaev:2023gaq,Agaev:2023rpj} as well, where we
evaluated parameters of various all-charmed scalar structures. Analyses were
done by employing SR method and diquark-antidiquark and hadronic molecule
models. It was found that the resonance $X(6200)$ is presumably $\eta
_{c}\eta _{c}$ molecule \cite{Agaev:2023ruu}. The structure $X(6600)$ was
considered in the diquark-antidiquark picture as a state built of
axial-vector components \cite{Agaev:2023wua}. Properties of resonance $%
X(6900)$, i.e., its mass and full width agree with data provided one models
it as a tetraquark composed of pseudoscalar diquarks and/or as a hadronic
molecule $\chi _{c0}\chi _{c0}$ \cite{Agaev:2023ruu,Agaev:2023gaq}.
Therefore, it was interpreted as an admixture of the diquark-antidiquark and
molecule-type states. The resonance $X(7300)$ was treated in Ref.\ \cite%
{Agaev:2023rpj} as a superposition of the molecule $\chi _{c1}\chi _{c1}$
and first radial excitation of $X(6600)$.

Recently the CMS collaboration informed about first measurements of $X$
resonances' quantum numbers \cite{CMS:2025fpt}. The parity $P$ and charge
conjugation $C$ symmetries of these structures were found to be $+1$. Their
spin $J$ \ is consistent with $J=2$, whereas $J=0$ and $J=1$ possibilities
were excluded at $95\%$ and $99\%$ confidence levels, respectively. The
collaboration also made suggestions about diquark-antidiquark nature of $%
X(6600)$ and considered structures $X(6900)$ and $X(7100)$ as its radial
excitations.

The tensor diquark-antidiquark state with internal composition $C\gamma
_{\mu }\otimes \gamma _{\nu }C+C\gamma _{\nu }\otimes \gamma _{\mu }C$ was
studied in our work \cite{Agaev:2026mif}. We calculated the mass $(6609\pm
50)~\mathrm{MeV}$ and decay width $(165\pm 23)~\mathrm{MeV}$ of this
tetraquark and compared them with available experimental data. We found that
within theoretical and experimental errors the mass agrees with experimental
data for the resonance $X(6600)$. But SR method led for the width of the
tetraquark to the prediction which is smaller than relevant CMS and ATLAS
data. At the same time, our theoretical limit $188~\mathrm{MeV}$ for the
width of this state is compatible with the lower border of the ATLAS's
data. Therefore, we interpreted the tensor tetraquark  $C\gamma _{\mu
}\otimes \gamma _{\nu }C+C\gamma _{\nu }\otimes \gamma _{\mu }C$  as a
tentative candidate to the resonance $X(6600)$.

In the current paper, we study the hadronic tensor molecule $\mathcal{M=}%
J/\psi J/\psi $ by computing its mass and decay width in the context of QCD
SR method \cite{Shifman:1978bx,Shifman:1978by}. The mass $m$ of $J/\psi
J/\psi $ is evaluated by utilizing the two-point sum rule approach. The
result for $m$ allows us to fix kinematically possible decay channels of the
molecule $\mathcal{M}$.

It turns out that the tensor molecule $\mathcal{M}$ decays to the meson pair
$J/\psi J/\psi $. This process is dominant decay channel of $\mathcal{M}$
because all constituent $c$ quarks (antiquarks) emerge in the final-state
mesons. But there exists alternative mechanism for transformation of the
molecule $\mathcal{M}$ to conventional mesons \cite%
{Becchi:2020mjz,Becchi:2020uvq,Agaev:2023ara}. This mechanism is connected
with annihilation of $c\overline{c}$ quarks in $\mathcal{M}$ into light $q%
\overline{q}$, $s\overline{s}$ quark pairs and production of $D_{(s)}^{(\ast
)+}D_{(s)}^{(\ast )-}$, $DD_{1}(2420),$ $D_{s}D_{s1}(2460)$ and $%
D_{(s)}^{(\ast )0}\overline{D}_{(s)}^{(\ast )0}$ mesons.

All decays of the molecule $\mathcal{M}$ are analyzed by utilizing methods
of the three-point sum rule approach. This is necessary to compute the form
factors $G(q^{2})$ and $g_{i}(q^{2})$ for relevant $\mathcal{M}$-meson-meson
vertices which at mass-shells give the strong couplings $G$ and $g_{i}$. In
their turn,  these couplings determine partial widths of channels under
consideration.

This paper is structured in the following manner: The spectroscopic
parameters of the tensor molecule $\mathcal{M}$ are computed in Sec.\ \ref%
{sec:Mass}. The dominant decay channel $\mathcal{M}\rightarrow J/\psi J/\psi
$ is considered in Sec.\ \ref{sec:Widths1}. The section \ref{sec:Widths2} is
devoted to analyses of six non-strange subdominant modes of $\mathcal{M}$.
Decays of $\mathcal{M}$ to charmed-strange mesons are explored in Sec.\ \ref%
{sec:Widths3}. In this section, we estimate also the full decay width of $%
\mathcal{M}$. We present analysis of obtained results and our final remarks
in Sec. \ref{sec:Conc}.


\section{Mass and current coupling of the molecule $\mathcal{M}$}

\label{sec:Mass}

The mass $m$ and current coupling $\Lambda $ of the tensor molecule $%
\mathcal{M}$ can be evaluated using the sum rules for these parameters. The
relevant SRs are extracted from analysis of the following correlation
function
\begin{equation}
\Pi _{\mu \nu \alpha \beta }(p)=i\int d^{4}xe^{ipx}\langle 0|\mathcal{T}%
\{J_{\mu \nu }(x)J_{\alpha \beta }^{\dag }(0)\}|0\rangle ,  \label{eq:CF1}
\end{equation}%
where $J_{\mu \nu }(x)$ is the interpolating current for the molecule $%
\mathcal{M}$.

The current $J_{\mu \nu }(x)$ is given by the expression
\begin{equation}
J_{\mu \nu }(x)=\overline{c}_{a}(x)\gamma _{\mu }c_{a}(x)c_{b}(x)\gamma
_{\nu }\overline{c}_{b}(x).
\end{equation}%
Here, $c(x)$ is the $c$-quark field with $a$ and $b$ being the color indices.

To determine the sum rules for $m$ and $\Lambda $, we express the
correlation function $\Pi _{\mu \nu \alpha \beta }(p)$ in terms of the
parameters of the molecule $\mathcal{M}$. The formula obtained by this way
is the phenomenological side $\Pi _{\mu \nu \alpha \beta }^{\mathrm{Phys}}(p)
$ of the SRs. For these purposes, we insert into $\Pi _{\mu \nu \alpha \beta
}(p)$ a complete set of intermediate states, carry out integration over $x$,
and find
\begin{eqnarray}
\Pi _{\mu \nu \alpha \beta }^{\mathrm{Phys}}(p) &=&\frac{\langle 0|J_{\mu
\nu }|\mathcal{M}(p,\epsilon )\rangle \langle \mathcal{M}(p,\epsilon
)|J_{\alpha \beta }^{\dag }|0\rangle }{m^{2}-p^{2}}  \notag \\
&&+\cdots .  \label{eq:CFphys}
\end{eqnarray}%
The polarization tensor of $\mathcal{M}$ is denoted above by $\epsilon
=\epsilon _{\mu \nu }(p)$. The term written down in Eq.\ (\ref{eq:CFphys}) is
a contribution of the ground-state molecule $\mathcal{M}$, whereas effects
of higher resonances and continuum states are presented by the dots.

The formula for $\Pi _{\mu \nu \alpha \beta }^{\mathrm{Phys}}(p)$ can be
written down by utilizing the matrix element
\begin{equation}
\langle 0|J_{\mu \nu }|\mathcal{M}(p,\epsilon (p)\rangle =\Lambda \epsilon
_{\mu \nu }(p).  \label{eq:ME1}
\end{equation}%
Having employed this element in the correlator $\Pi _{\mu \nu \alpha \beta
}^{\mathrm{Phys}}(p)$ and carried out necessary manipulations, we get
\begin{eqnarray}
\Pi _{\mu \nu \alpha \beta }^{\mathrm{Phys}}(p) &=&\frac{\Lambda ^{2}}{%
m^{2}-p^{2}}\left\{ \frac{1}{2}\left( g_{\mu \alpha }g_{\nu \beta }+g_{\mu
\beta }g_{\nu \alpha }\right) \right.  \notag \\
&&\left. +\text{ other components}\right\} +\cdots .  \label{eq:Phys2}
\end{eqnarray}%
The function $\Pi _{\mu \nu \alpha \beta }^{\mathrm{Phys}}(p)$ contains
different Lorentz structures. Because the term $\sim (g_{\mu \alpha }g_{\nu
\beta }+g_{\mu \beta }g_{\nu \alpha })$ comes from only of a spin-$2$
particle, we use it in our analysis and label corresponding invariant
amplitude by $\Pi ^{\mathrm{Phys}}(p^{2})$.

To determine QCD side of the sum rules $\Pi _{\mu \nu \alpha \beta }^{%
\mathrm{OPE}}(p)$ we insert $J_{\mu \nu }(x)$ into $\Pi _{\mu \nu \alpha
\beta }(p)$ and contract quark fields. We get
\begin{eqnarray}
&&\Pi _{\mu \nu \alpha \beta }^{\mathrm{OPE}}(p)=2i\int d^{4}xe^{ipx}\left\{
\mathrm{Tr}\left[ \gamma _{\mu }S_{c}^{aa^{\prime }}(x)\gamma _{\alpha
}S_{c}^{a^{\prime }a}(-x)\right] \right.  \notag \\
&&\times \mathrm{Tr}\left[ \gamma _{\nu }S_{c}^{bb^{\prime }}(x)\gamma
_{\beta }S_{c}^{b^{\prime }b}(-x)\right] -\mathrm{Tr}\left[ \gamma _{\mu
}S_{c}^{aa^{\prime }}(x)\gamma _{\alpha }\right.  \notag \\
&&\left. \left. \times S_{c}^{a^{\prime }b}(-x)\gamma _{\nu
}S_{c}^{bb^{\prime }}(x)\gamma _{\beta }S_{c}^{b^{\prime }a}(-x)\right]
\right\} ,
\end{eqnarray}%
where $S_{c}(x)$ is $c-$quark propagator \cite{Agaev:2020zad}.

The correlator $\Pi _{\mu \nu \alpha \beta }^{\mathrm{OPE}}(p)$ has to be
computed using operator product expansion ($\mathrm{OPE}$) with some
accuracy. Having extracted a contribution in $\Pi _{\mu \nu \alpha \beta }^{%
\mathrm{OPE}}(p)$ proportional to $(g_{\mu \alpha }g_{\nu \beta }+g_{\mu
\beta }g_{\nu \alpha })$ and labeled by $\Pi ^{\mathrm{OPE}}(p^{2})$ the
relevant amplitude one can determine the required SRs. To this end, one
equates the amplitudes $\Pi ^{\mathrm{Phys}}(p^{2})$ and $\Pi ^{\mathrm{OPE}%
}(p^{2})$ and performs standard operations of SR method. In other words, one
applies the Borel transformation which suppresses contributions of higher
resonances and continuum states. Afterwards, using the quark-hadron duality
assumption, one subtracts these contributions from QCD side of the SR
equality. After these manipulations $\Pi ^{\mathrm{OPE}}(p^{2})$ becomes
equal to $\Pi (M^{2},s_{0})$ and depends on the Borel and continuum
subtraction parameters $M^{2}$ and $s_{0}$. The SRs for the mass $m$ and
current coupling $\Lambda $ read
\begin{equation}
m^{2}=\frac{\Pi ^{\prime }(M^{2},s_{0})}{\Pi (M^{2},s_{0})},  \label{eq:Mass}
\end{equation}%
and
\begin{equation}
\Lambda ^{2}=e^{m^{2}/M^{2}}\Pi (M^{2},s_{0}),  \label{eq:Coupl}
\end{equation}%
where $\Pi ^{\prime }(M^{2},s_{0})=d\Pi (M^{2},s_{0})/d(-1/M^{2})$.

The transformed amplitude $\Pi (M^{2},s_{0})$ is given by the formula%
\begin{equation}
\Pi (M^{2},s_{0})=\int_{16m_{c}^{2}}^{s_{0}}ds\rho ^{\mathrm{OPE}%
}(s)e^{-s/M^{2}}+\Pi (M^{2}).  \label{eq:CorrF}
\end{equation}%
We compute it by taking into account dimension-$4$ terms $\sim \langle
\alpha _{s}G^{2}/\pi \rangle $. In Eq.\ (\ref{eq:CorrF}) $\rho ^{\mathrm{OPE}%
}(s)$ is the spectral density which amounts to the imaginary part of the
amplitude $\Pi ^{\mathrm{OPE}}(p^{2})$. The contribution $\Pi (M^{2})$ is
obtained directly from $\Pi ^{\mathrm{OPE}}(p^{2})$ and contains terms
absent in $\rho ^{\mathrm{OPE}}(s)$.

For numerical analysis we should determine parameters in the relevant SRs.
For the mass $m_{c}$ of $c$ quark and gluon condensate $\langle \alpha
_{s}G^{2}/\pi \rangle $ which are universal entries, we utilize%
\begin{eqnarray}
&&m_{c}=(1.2730\pm 0.0046)~\mathrm{GeV},  \notag \\
&&\langle \alpha _{s}G^{2}/\pi \rangle =(0.012\pm 0.004)~\mathrm{GeV}^{4}.
\end{eqnarray}

The parameters $M^{2}$ and $s_{0}$ should be chosen for each process under
investigation and have to meet standard constraints of SR analysis. These
constraints imply prevalence of the pole contribution ($\mathrm{PC}$) in
extracted parameters, convergence of $\mathrm{OPE}$ and minimal dependence
of $m$ and $\Lambda $ on parameters $M^{2}$ and $s_{0}$: These constants are
important for the credibility of SR results. Therefore, we require
fulfillment of $\mathrm{PC}\geq 0.5$, where
\begin{equation}
\mathrm{PC}=\frac{\Pi (M^{2},s_{0})}{\Pi (M^{2},\infty )},  \label{eq:PC}
\end{equation}

Because $\Pi (M^{2},s_{0})$ contains the perturbative and dimension-$4$
contribution $\Pi ^{\mathrm{Dim4}}(M^{2},s_{0})$, to ensure convergence of $%
\mathrm{OPE}$, we impose the constraint $|\Pi ^{\mathrm{Dim4}%
}(M^{2},s_{0})|\leq 0.05|\Pi (M^{2},s_{0})|$. Note that these two
restrictions allow us to find the maximal and minimal values of $M^{2}$,
respectively.

Calculations prove that regions for $M^{2}$ and $s_{0}$
\begin{equation}
M^{2}\in \lbrack 4.5,5.5]~\mathrm{GeV}^{2},\ s_{0}\in \lbrack 45,46]~\mathrm{%
GeV}^{2},  \label{eq:Wind1}
\end{equation}%
satisfy all necessary constraints. In fact, at $6.5~\mathrm{GeV}^{2}$ and $%
5.5~\mathrm{GeV}^{2}$ the pole contribution on the average in $s_{0}$  is
equal to $\mathrm{PC}\approx 0.49$ and $\mathrm{PC}\approx 0.76$ ,
respectively. The term $|\Pi ^{\mathrm{Dim4}}(M^{2},s_{0})|$ at $M^{2}=5~%
\mathrm{GeV}^{2}$ is around $2\%$ of the amplitude $\Pi (M^{2},s_{0})$.

We evaluate $m$ and $\Lambda $ in the regions Eq.\ (\ref{eq:Wind1}) and
determine their average values
\begin{eqnarray}
&&m=(6290\pm 50)~\mathrm{MeV},  \notag \\
&&\Lambda =(1.85\pm 0.15)\times 10^{-1}~\mathrm{GeV}^{5}.  \label{eq:Result1}
\end{eqnarray}

The predictions in Eq.\ (\ref{eq:Result1}) are equivalent to SR results at
the point $M^{2}=5~\mathrm{GeV}^{2}$ and $s_{0}=45.5~\mathrm{GeV}^{2}$,
where $\mathrm{PC}\approx 0.61$. This ensures dominance of $\mathrm{PC}$ in
extracted parameters $m$ and $\Lambda $.
\begin{widetext}

\begin{figure}[htbp]
\begin{center}
\includegraphics[totalheight=6cm,width=8cm]{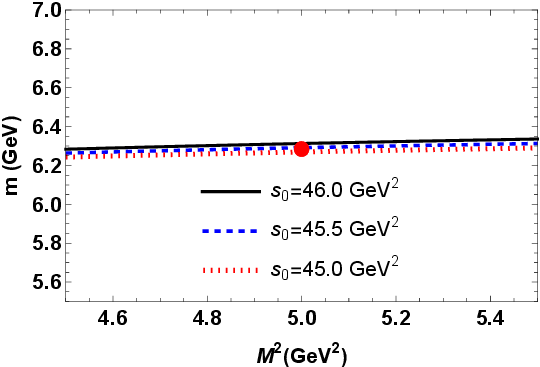}
\includegraphics[totalheight=6cm,width=8cm]{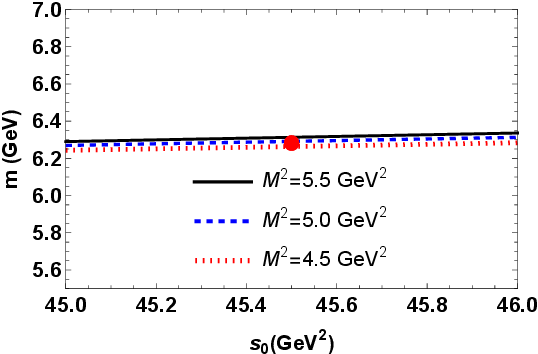}
\end{center}
\caption{The mass $m$ as a function of the parameters $M^{2}$ (left), and $s_0$ (right).}
\label{fig:Mass}
\end{figure}

\end{widetext}

Ambiguities in Eq.\ (\ref{eq:Result1}) are connected with those in $M^{2}$
and $s_{0}$: The quark mass $m_{c}$ and condensate $\langle \alpha
_{s}G^{2}/\pi \rangle $ almost do not generate sizeable errors.
Uncertainties of the mass $m$ are equal to $\pm 0.79\%$ of the central
value, whereas for $\Lambda $ they amount to $\pm 8.11\%$. All these
theoretical errors are inside of usual limits of SR analysis proving
validity of the obtained results. Dependencies of the mass $m$ on the Borel
and continuum subtraction parameters are depicted in Fig.\ \ref{fig:Mass}.


\section{Dominant decay channel of the molecule $\mathcal{M}$}

\label{sec:Widths1}


Result for the mass $m$ of the tensor molecule $\mathcal{M}$ and its quantum
numbers $J^{\mathrm{PC}}=2^{++}$ allow us to find its dominant decay
channel. It is easy to see that decay to a pair of the vector $J/\psi $
mesons is the kinematically permitted mode of $\mathcal{M}$. In fact,
two-meson threshold for this process $6194~$\textrm{M}$\mathrm{eV}$ is below
the mass of the molecule $\mathcal{M}$. Even in lower limit of the mass $%
m=6240~\mathrm{MeV}$ the molecule $\mathcal{M}$ easily decays through this
channel.

To calculate the partial width of the decay $\mathcal{M}\rightarrow J/\psi
J/\psi $ we have to find the strong coupling $G$ that describes the
interaction of particles at the vertex $\mathcal{M}J/\psi J/\psi $ . The
analysis of the three-point correlation function
\begin{eqnarray}
\Pi _{\mu \nu \alpha \beta }(p,p^{\prime }) &=&i^{2}\int
d^{4}xd^{4}ye^{ip^{\prime }y}e^{-ipx}\langle 0|\mathcal{T}\{J_{\mu }^{J/\psi
}(y)  \notag \\
&&\times J_{\nu }^{J/\psi }(0)J_{\alpha \beta }^{\dagger }(x)\}|0\rangle ,
\label{eq:CF1a}
\end{eqnarray}%
will allow us to derive SR for the form factor $G(q^{2})$ which at the mass
shell $q^{2}=m_{J/\psi }^{2}$ is equal to the strong coupling $G$. In Eq.\ (%
\ref{eq:CF1a}) $J_{\mu }^{J/\psi }(x)$ is the current which interpolate the
vector charmonium $J/\psi $ $\ $%
\begin{equation}
J_{\mu }^{J/\psi }(x)=\overline{c}_{i}(x)\gamma _{\mu }c_{i}(x),
\end{equation}%
where $i$ is the color index.

The correlation function $\Pi _{\mu \nu \alpha \beta }(p,p^{\prime })$
expressed using parameters of particles $\mathcal{M}$ and $J/\psi $ gives
the phenomenological side $\Pi _{\mu \nu \alpha \beta }^{\mathrm{Phys}%
}(p,p^{\prime })$ of SR. Having taken into account contribution of the
ground-level particles, we get%
\begin{eqnarray}
&&\Pi _{\mu \nu \alpha \beta }^{\mathrm{Phys}}(p,p^{\prime })=\frac{\langle
0|J_{\mu }^{J/\psi }|J/\psi (p^{\prime },\varepsilon _{1})\rangle }{%
p^{\prime 2}-m_{J/\psi }^{2}}\frac{\langle 0|J_{\nu }^{J/\psi }|J/\psi
(q,\varepsilon _{2})\rangle }{q^{2}-m_{J/\psi }^{2}}  \notag \\
&&\times \langle J/\psi (p^{\prime },\varepsilon _{1})J/\psi (q,\varepsilon
_{2})|\mathcal{M}(p,\epsilon )\rangle \frac{\langle \mathcal{M}%
(p,\varepsilon )|J_{\alpha \beta }^{\dagger }|0\rangle }{p^{2}-m^{2}}  \notag
\\
&&+\cdots .  \label{eq:TP1}
\end{eqnarray}%
Here, $m_{J/\psi }$ is the mass $(3096.900\pm 0.006)~\mathrm{MeV}$ of the
meson $J/\psi $ \cite{PDG:2024}, whereas the polarization vectors of $J/\psi
(p^{\prime },\varepsilon _{1})$ and $J/\psi (q,\varepsilon _{2})$ are
denoted by $\varepsilon _{1}$ and $\varepsilon _{2}$, respectively.

Equation (\ref{eq:TP1}) can be transformed into a more convenient form. With
this in mind, we introduce the matrix elements%
\begin{eqnarray}
\langle 0|J_{\mu }^{J/\psi }|J/\psi (p^{\prime },\varepsilon _{1})\rangle
&=&f_{J/\psi }m_{J/\psi }\varepsilon _{1\mu }(p^{\prime }),  \notag \\
\langle 0|J_{\nu }^{J/\psi }|J/\psi (q,\varepsilon _{2})\rangle &=&f_{J/\psi
}m_{J/\psi }\varepsilon _{2\nu }(q),  \label{eq:C2}
\end{eqnarray}%
with $f_{J/\psi }=(411\pm 7)~\mathrm{MeV}$ being the decay constant of $%
J/\psi $ \cite{Lakhina:2006vg}.

The vertex matrix element $\langle J/\psi (p^{\prime },\varepsilon
_{1})J/\psi (q,\varepsilon _{2})|\mathcal{M}(p,\epsilon )\rangle $ can be
expressing by employing the parameters of the molecule $\mathcal{M}$ and
mesons $J/\psi $ in the following form \cite{Agaev:2024pil}
\begin{eqnarray}
&&\langle J/\psi (p^{\prime },\varepsilon _{1})J/\psi (q,\varepsilon _{2})|%
\mathcal{M}(p,\epsilon )\rangle =G(q^{2})\epsilon _{\tau \rho }\left[
(\varepsilon _{1}^{\ast }\cdot q)\right.  \notag \\
&&\times \varepsilon _{2}^{\tau \ast }p^{\prime \rho }+(\varepsilon
_{2}^{\ast }\cdot p^{\prime })\varepsilon _{1}^{\ast \tau }q^{\rho
}-(p^{\prime }\cdot q)\varepsilon _{1}^{\tau \ast }\varepsilon _{2}^{\rho
\ast }  \notag \\
&&\left. -(\varepsilon _{1}^{\ast }\cdot \varepsilon _{2}^{\ast })p^{\prime
\tau }q^{\rho }\right] .  \label{eq:TVV}
\end{eqnarray}%
Then, for $\Pi _{\mu \nu \alpha \beta }^{\mathrm{Phys}}(p,p^{\prime })$ we
find
\begin{eqnarray}
&&\Pi _{\mu \nu \alpha \beta }^{\mathrm{Phys}}(p,p^{\prime })=G(q^{2})\frac{%
\Lambda f_{J/\psi }^{2}m_{J/\psi }^{2}}{\left( p^{2}-m^{2}\right) (p^{\prime
2}-m_{J/\psi }^{2})}  \notag \\
&&\times \frac{1}{(q^{2}-m_{J/\psi }^{2})}\left[ p_{\beta }^{\prime
}p_{\alpha }^{\prime }g_{\mu \nu }+\frac{1}{2}p_{\mu }p_{\alpha }^{\prime
}g_{\beta \nu }\right.  \notag \\
&&\left. +\frac{1}{2m^{2}}p_{\beta }p_{\nu }p_{\mu }^{\prime }p_{\alpha
}^{\prime }+\text{ other terms}\right] +\cdots .
\end{eqnarray}

The correlation function $\Pi _{\mu \nu \alpha \beta }^{\mathrm{OPE}%
}(p,p^{\prime })$ is given by the expression
\begin{eqnarray}
&&\Pi _{\mu \nu \alpha \beta }^{\mathrm{OPE}}(p,p^{\prime })=i^{2}\int
d^{4}xd^{4}ye^{ip^{\prime }y}e^{-ipx}\left\{ \mathrm{Tr}\left[ \gamma _{\mu
}S_{c}^{ia}(y-x)\right. \right.  \notag \\
&&\left. \times \gamma _{\alpha }S_{c}^{ai}(x-y)\right] \mathrm{Tr}\left[
\gamma _{\nu }S_{c}^{jb}(-x)\gamma _{\beta }S_{c}^{bj}(x)\right]  \notag \\
&&-\mathrm{Tr}\left[ \gamma _{\mu }S_{c}^{ib}(y-x)\gamma _{\beta
}S_{c}^{bi}(x-y)\right] \mathrm{Tr}\left[ \gamma _{\nu }S_{c}^{ja}(-x)\gamma
_{\alpha }S_{c}^{aj}(x)\right]  \notag \\
&&+\mathrm{Tr}\left[ \gamma _{\mu }S_{c}^{ib}(y-x)\gamma _{\beta
}S_{c}^{bj}(x)\gamma _{\nu }S_{c}^{ja}(-x)\gamma _{\alpha }S_{c}^{ai}(x-y)%
\right]  \notag \\
&&\left. -\mathrm{Tr}\left[ \gamma _{\mu }S_{c}^{ia}(y-x)\gamma _{\alpha
}S_{c}^{aj}(x)\gamma _{\nu }S_{c}^{jb}(-x)\gamma _{\beta }S_{c}^{bi}(x-y)%
\right] \right\} .  \notag \\
&&
\end{eqnarray}%
We use the amplitudes $\Pi ^{\mathrm{Phys}}(p^{2},p^{\prime 2},q^{2})$ and $%
\Pi ^{\mathrm{OPE}}(p^{2},p^{\prime 2},q^{2})$ that correspond to terms $%
\sim p_{\beta }p_{\nu }p_{\mu }^{\prime }p_{\alpha }^{\prime }$ in these
correlators, and derive SR for $G(q^{2})$. Usual manipulations yield
\begin{equation}
G(q^{2})=\frac{2m^{2}(q^{2}-m_{J/\psi }^{2})}{\Lambda f_{J/\psi
}^{2}m_{J/\psi }^{2}}e^{m^{2}/M_{1}^{2}}e^{m_{J/\psi }^{2}/M_{2}^{2}}\Pi (%
\mathbf{M}^{2},\mathbf{s}_{0},q^{2}).  \label{eq:SRG}
\end{equation}%
In Eq.\ (\ref{eq:SRG}), $\Pi (\mathbf{M}^{2},\mathbf{s}_{0},q^{2})$ is the
function $\Pi ^{\mathrm{OPE}}(p^{2},p^{\prime 2},q^{2})$ after the Borel
transformations and continuum subtractions. It depends on the parameters $%
\mathbf{M}^{2}=(M_{1}^{2},M_{2}^{2})$ and $\mathbf{s}_{0}=(s_{0},s_{0}^{%
\prime })$. The pair $(M_{1}^{2},s_{0})$ corresponds to the channel of the
molecule $\mathcal{M}$, whereas $(M_{2}^{2},s_{0}^{\prime })$ is related to $%
J/\psi $ channel. The function $\Pi (\mathbf{M}^{2},\mathbf{s}_{0},q^{2})$
is determined as%
\begin{eqnarray}
&&\Pi (\mathbf{M}^{2},\mathbf{s}_{0},q^{2})=\int_{16m_{c}^{2}}^{s_{0}}ds%
\int_{4m_{c}^{2}}^{s_{0}^{\prime }}ds^{\prime }\rho (s,s^{\prime },q^{2})
\notag \\
&&\times e^{-s/M_{1}^{2}-s^{\prime }/M_{2}^{2}}.  \label{eq:CorrF1}
\end{eqnarray}

Constraints on $\mathbf{M}^{2}$ and $\mathbf{s}_{0}$ are standard for SR
investigations and have been detailed in the previous section. Our
calculations prove that working windows Eq.\ (\ref{eq:Wind1}) for the
parameters $(M_{1}^{2},s_{0})$ and
\begin{equation}
M_{2}^{2}\in \lbrack 4,5]~\mathrm{GeV}^{2},\ s_{0}^{\prime }\in \lbrack
12,13]~\mathrm{GeV}^{2},  \label{eq:Wind3}
\end{equation}%
for $(M_{2}^{2},s_{0}^{\prime })$ meet these conditions.

\begin{figure}[h]
\includegraphics[width=8.5cm]{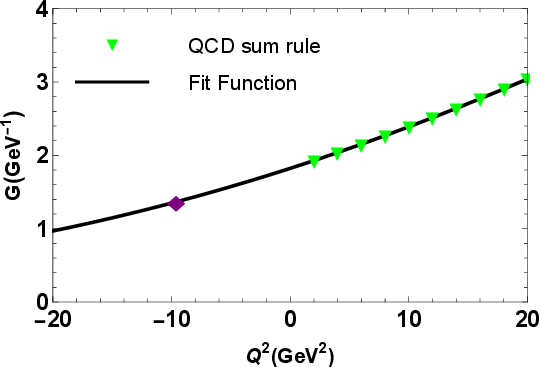}
\caption{SR data and fit function $\mathcal{Z}(Q^{2})$ for the strong
coupling $G$. The diamond marks the point $\mathcal{Z}(-m_{J/\protect\psi%
}^{2})$. }
\label{fig:Fit}
\end{figure}

The sum rule for the form factor $G(q^{2})$ is applicable in the region $%
q^{2}<0$. But $G(q^{2})$ gives the coupling $G$ at the mass shell $%
q^{2}=m_{J/\psi }^{2}$. For that reason, we employ the function $G(Q^{2})$
where $Q^{2}=-q^{2}$ and apply it in following studies. SR predictions for $%
G(Q^{2})$ in the interval $Q^{2}=2-20~\mathrm{GeV}^{2}$ are shown in Fig.\ %
\ref{fig:Fit}.

It has been emphasized above that $G$ has to be estimated at $%
q^{2}=m_{J/\psi }^{2}$, i.e., at $Q^{2}=-m_{J/\psi }^{2}$. But at that point
one cannot use directly the SR method. To avoid this obstacle, we employ the
function $\mathcal{Z}(Q^{2})$ which at $Q^{2}>0$ amounts to SR data $%
G(Q^{2}) $, but can be extrapolated to the region $Q^{2}<0$. For these
purposes, we employ
\begin{equation}
\mathcal{Z}_{i}(Q^{2})=\mathcal{Z}_{i}^{0}\mathrm{\exp }\left[ z_{i}^{1}%
\frac{Q^{2}}{m^{2}}+z_{i}^{2}\left( \frac{Q^{2}}{m^{2}}\right) ^{2}\right] ,
\label{eq:FitF}
\end{equation}%
where $\mathcal{Z}_{i}^{0}$, $z_{i}^{1}$, and $z_{i}^{2}$ are fitting
parameters. Having compared SR data and Eq.\ (\ref{eq:FitF}), it is easy to
fix
\begin{equation}
\mathcal{Z}^{0}=1.823~\mathrm{GeV}^{-1},\ z^{1}=1.131,\text{ }z^{2}=-0.234.
\label{eq:FF1}
\end{equation}%
The function $\mathcal{Z}(Q^{2})$ is drawn in Fig.\ \ref{fig:Fit}, where
agreement with the sum rule's data is evident. For $G$, one gets
\begin{equation}
G\equiv \mathcal{Z}(-m_{J/\psi }^{2})=(1.37\pm 0.26)\ \mathrm{GeV}^{-1}.
\label{eq:g1}
\end{equation}

Partial width of the process $\mathcal{M}\rightarrow J/\psi J/\psi $ is
determined by the expression%
\begin{equation}
\Gamma \left[ \mathcal{M}\rightarrow J/\psi J/\psi \right] =\frac{%
G^{2}\lambda }{2\cdot 80\pi m^{2}}(m^{4}-3m^{2}m_{J/\psi }^{2}+6m_{J/\psi
}^{4}).  \label{eq:PDw2}
\end{equation}%
In Eq.\ (\ref{eq:PDw2}) $\lambda =\lambda (m,m_{J/\psi },m_{J/\psi })$ is
defined as
\begin{equation}
\lambda (x,y,z)=\frac{\sqrt{%
x^{4}+y^{4}+z^{4}-2(x^{2}y^{2}+x^{2}z^{2}+y^{2}z^{2})}}{2x}.
\end{equation}%
As a result, we find
\begin{equation}
\Gamma \left[ \mathcal{M}\rightarrow J/\psi J/\psi \right] =(50.2\pm 18.3)~%
\mathrm{MeV}.  \label{eq:DW2}
\end{equation}

\section{Subdominant modes of $\mathcal{M}$}

\label{sec:Widths2}


The molecule $\mathcal{M}$ transforms to conventional particles through
annihilation of $c\overline{c}$ quarks to $q\overline{q}$, $s\overline{s}$
pairs \cite{Becchi:2020mjz,Becchi:2020uvq,Agaev:2023ara} and creation of $%
D_{(s)}D_{(s)}$ and $D_{(s)}D_{1(s)}$ mesons with appropriate quantum
numbers. Here, we consider the processes $\mathcal{M}\rightarrow D^{\ast
+}D^{\ast -},\ D^{\ast 0}\overline{D}^{\ast 0}$, $D^{+}D_{1}^{-}$, $D^{0}%
\overline{D}_{1}^{0}$, $D^{-}D_{1}^{+}$,$\ \overline{D}^{0}D_{1}^{0}$, $%
D^{+}D^{-}$,$\ D^{0}\overline{D}^{0}$, as well as decays to $D_{s}^{\ast
+}D_{s}^{\ast -}$, $D_{s}^{+}D_{s1}^{-}$, $D_{s}^{-}D_{s1}^{+}$, and $%
D_{s}^{+}D_{s}^{-}$ meson pairs.


\subsection{Processes $\mathcal{M}\rightarrow D^{\ast +}D^{\ast -}$ and $%
D^{\ast 0}\overline{D}^{\ast 0}$}


In this subsection, we consider decays $\mathcal{M}\rightarrow D^{\ast
+}D^{\ast -}$ and $D^{\ast 0}\overline{D}^{\ast 0}$. Note that the
correlation function of these decays differ from each other only by
propagators of $d$ and $u$ quarks. Because we adopt the approximation $%
m_{u}=m_{d}=0$, and also neglect small numerical differences in the masses
of charged and neutral $D$ mesons, the decays $\mathcal{M}\rightarrow
D^{\ast +}D^{\ast -}$ and $D^{\ast 0}\overline{D}^{\ast 0}$ have the same
partial widths.

Therefore, let us concentrate on the channel $\mathcal{M}\rightarrow D^{\ast
+}D^{\ast -}$. To determine the strong coupling $g_{4}$ at the$\mathcal{M}$%
-meson-meson vertex $\mathcal{M}D^{\ast +}D^{\ast -}$, we study the
correlation function%
\begin{eqnarray}
\widetilde{\Pi }_{\mu \nu \alpha \beta }(p,p^{\prime }) &=&i^{2}\int
d^{4}xd^{4}ye^{ip^{\prime }y}e^{-ipx}\langle 0|\mathcal{T}\{J_{\mu
}^{D^{\ast +}}(y)  \notag \\
&&\times J_{\nu }^{D^{\ast -}}(0)J_{\alpha \beta }^{\dagger }(x)\}|0\rangle ,
\label{eq:CF1A}
\end{eqnarray}%
where $J_{\mu }^{D^{\ast +}}(x)$ and $J_{\nu }^{D^{\ast -}}(x)$ are the
interpolating currents of the mesons $D^{\ast +}$ and $D^{\ast -}$
\begin{equation}
J_{\mu }^{D^{\ast +}}(x)=\overline{d}_{i}(x)\gamma _{\mu }c_{i}(x),\ J_{\nu
}^{D^{\ast -}}(x)=\overline{c}_{j}(x)\gamma _{\nu }d_{j}(x).\text{ }
\label{eq:CRB}
\end{equation}

In terms of the matrix elements of the particles $\mathcal{M}$, $D^{\ast +}$%
, and $D^{\ast -}$ the correlator $\widetilde{\Pi }_{\mu \nu \alpha \beta
}(p,p^{\prime })$ is
\begin{eqnarray}
&&\widetilde{\Pi }_{\mu \nu \alpha \beta }^{\mathrm{Phys}}(p,p^{\prime })=%
\frac{\langle 0|J_{\mu }^{D^{\ast +}}|D^{\ast +}(p^{\prime },\varepsilon
_{1})\rangle }{p^{\prime 2}-m_{D^{\ast }}^{2}}\frac{\langle 0|J_{\nu
}^{D^{\ast -}}|D^{\ast -}(q,\varepsilon _{2})\rangle }{q^{2}-m_{D^{\ast
}}^{2}}  \notag  \label{eq:CF2} \\
&&\times \langle D^{\ast +}(p^{\prime },\varepsilon _{1})D^{\ast
-}(q,\varepsilon _{2})|\mathcal{M}(p,\epsilon )\rangle \frac{\langle
\mathcal{M}(p,\epsilon )|I_{\alpha \beta }^{\dagger }|0\rangle }{p^{2}-m^{2}}
\notag \\
&&+\cdots ,
\end{eqnarray}%
where $m_{D^{\ast }}$ is the mass of the mesons $D^{\ast \pm }$ that equals
to $(2010.26\pm 0.05)~\mathrm{MeV}$, while $\varepsilon _{1\mu }$ and $%
\varepsilon _{2\nu }$ are their polarization vectors, respectively.

The correlator $\widetilde{\Pi }_{\mu \nu \alpha \beta }^{\mathrm{Phys}%
}(p,p^{\prime })$ is obtained using matrix elements
\begin{eqnarray}
\langle 0|J_{\mu }^{D^{\ast +}}|D^{\ast +}(p^{\prime },\varepsilon
_{1})\rangle &=&f_{D^{\ast }}m_{D^{\ast }}\varepsilon _{1\mu }(p^{\prime }),
\notag \\
\langle 0|J_{\nu }^{D^{\ast -}}|D^{\ast -}(q,\varepsilon _{2})\rangle
&=&f_{D^{\ast }}m_{D^{\ast }}\varepsilon _{2\nu }(q),  \label{eq:ME2B}
\end{eqnarray}%
with $f_{D^{\ast }}=(252.2\pm 22.66)~\mathrm{MeV}$ being the decay constants
of the mesons $D^{\ast \pm }$ \cite{Lucha:2014spa}. The vertex $\langle
D^{\ast +}(p^{\prime },\varepsilon _{1})D^{\ast -}(q,\varepsilon _{2})|%
\mathcal{M}(p,\epsilon )\rangle $ is given by Eq.\ (\ref{eq:TVV}).

A sum rule for the form factor $g_{1}(q^{2})$ is obtained by employing the
amplitude $\Pi _{1}^{\mathrm{Phys}}(p^{2},p^{\prime 2},q^{2})$ that
corresponds in $\widetilde{\Pi }_{\mu \nu \alpha \beta }^{\mathrm{Phys}%
}(p,p^{\prime })$ to the term $\sim g_{\mu \nu }g_{\alpha \beta }$. The
function $\widetilde{\Pi }_{\mu \nu \alpha \beta }(p,p^{\prime })$
calculated in terms of quark propagators equals to
\begin{eqnarray}
&&\widetilde{\Pi }_{\mu \nu \alpha \beta }^{\mathrm{OPE}}(p,p^{\prime })=%
\frac{2}{3}g_{\alpha \beta }\int d^{4}xd^{4}ye^{ip^{\prime
}y}e^{-ipx}\langle \overline{c}c\rangle   \notag \\
&&\times \mathrm{Tr}\left[ \gamma _{\mu }S_{d}^{ij}(y)\gamma _{\nu
}S_{c}^{ja}(-x){}S_{c}^{ai}(x-y)\right] ,  \label{eq:QCDsideA}
\end{eqnarray}%
where $S_{d}(x)$ is the $d$ quark's propagator \cite{Agaev:2020zad} and $%
\langle \overline{c}c\rangle $ vacuum matrix element of $\ \overline{c}c$.
We label by $\Pi _{1}^{\mathrm{OPE}}(p^{2},p^{\prime 2},q^{2})$ the
amplitude that corresponds in $\widetilde{\Pi }_{\mu \nu \alpha \beta }^{%
\mathrm{OPE}}(p,p^{\prime })$ to the same structure $\sim g_{\mu \nu
}g_{\alpha \beta }$.

In what follows, we utilize the relation
\begin{equation}
\langle \overline{c}c\rangle \approx -\frac{1}{12m_{c}}\langle \frac{\alpha
_{s}G^{2}}{\pi }\rangle  \label{eq:Conden}
\end{equation}%
between the condensates obtained in Ref.\ \cite{Shifman:1978bx}.

\begin{figure}[h]
\includegraphics[width=8.5cm]{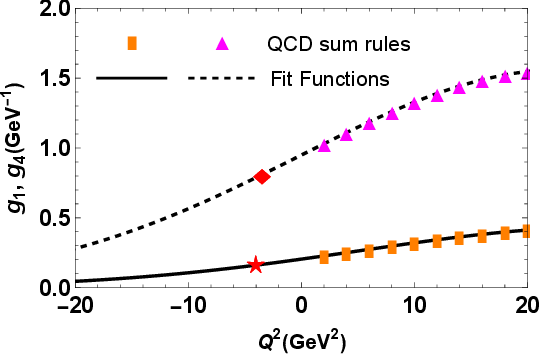}
\caption{QCD data and extrapolating functions $\mathcal{Z}_1(Q^{2})$ (solid
line) and $\mathcal{Z}_4(Q^{2})$ (dashed line) for the strong couplings $g_1$
and $g_4$, respectively. The star and circle mark the points $\mathcal{Z}%
_1(-m_{D^{\ast}}^{2})$ and $\mathcal{Z}_4(-m_{D}^{2})$. }
\label{fig:Fit1}
\end{figure}

SR for the form factor $g_{1}(q^{2})$ reads%
\begin{eqnarray}
&&g_{1}(q^{2})=\frac{12m^{2}(q^{2}-m_{D^{\ast }}^{2})}{\Lambda f_{D^{\ast
}}^{2}m_{D^{\ast }}^{2}[3m^{4}-4m^{2}(m_{D}^{2}+q^{2})+(m_{D}^{2}-q^{2})^{2}]%
}  \notag \\
&&\times e^{m^{2}/M_{1}^{2}}e^{m_{D^{\ast }}^{2}/M_{2}^{2}}\Pi _{1}(\mathbf{M%
}^{2},\mathbf{s}_{0},q^{2}).
\end{eqnarray}%
For the $D^{\ast }$ meson's channel, we employ the windows%
\begin{equation}
M_{2}^{2}\in \lbrack 3,5]~\mathrm{GeV}^{2},\ s_{0}^{\prime }\in \lbrack 6,8]~%
\mathrm{GeV}^{2}.  \label{eq:Wind2}
\end{equation}%
To find $g_{1}$ we employ SR data for $Q^{2}=2-20~\mathrm{GeV}^{2}$ and
extrapolating function with parameters $\mathcal{Z}_{1}^{0}=0.205~\mathrm{GeV%
}^{-1}$, $z_{1}^{1}=2.173$, and $z_{1}^{2}=-1.558$. The function $\mathcal{Z}%
_{1}(Q^{2})$ and corresponding SR results are shown in Fig.\ \ref{fig:Fit1}.

The coupling $g_{1}$ is evaluated at $q^{2}=m_{D^{\ast }}^{2}$ and is equal
to
\begin{equation}
g_{1}\equiv \mathcal{Z}_{1}(-m_{D^{\ast }}^{2})=(1.62\pm 0.31)\times
10^{-1}\ \mathrm{GeV}^{-1}.  \label{eq:G1}
\end{equation}%
The width of the process $\mathcal{M}\rightarrow D^{\ast +}D^{\ast -}$ is
\begin{equation}
\Gamma \left[ \mathcal{M}\rightarrow D^{\ast +}D^{\ast -}\right] =(7.5\pm
2.1)~\mathrm{MeV}.
\end{equation}

A difference between decays $\mathcal{M}\rightarrow D^{\ast +}D^{\ast -}$
and $\mathcal{M}\rightarrow D^{\ast 0}\overline{D}^{\ast 0}$ appears owing
to masses of the final-state mesons. With reasonable accuracy we adopt $%
\Gamma \left[ \mathcal{M}\rightarrow D^{\ast +}D^{\ast -}\right] \approx
\Gamma \left[ \mathcal{M}\rightarrow D^{\ast 0}\overline{D}^{\ast 0}\right] $%
.

\subsection{Decays $\mathcal{M}\rightarrow DD_{1}(2420)$}


The channels considering here are decays of $\mathcal{M}$ to pairs of $%
D^{+}D_{1}^{-}$, $D^{0}\overline{D}_{1}^{0}$, $D^{-}D_{1}^{+}$ and$\
\overline{D}^{0}D_{1}^{0}$ mesons. Evidently all these processes are allowed
decay modes of the molecule $\mathcal{M}$. We collect them into two groups $%
\mathcal{M}\rightarrow D^{+}D_{1}^{-}$, $D^{0}\overline{D}_{1}^{0}$ and $%
\mathcal{M}\rightarrow D^{-}D_{1}^{+}$, $\overline{D}^{0}D_{1}^{0}$ because
inside of groups processes are connected with each other by $d\rightarrow u$
replacement. As it has been explained above we treat such channels as decays
with the same widths. These means that in our approximation $\Gamma \left[
\mathcal{M}\rightarrow D^{+}D_{1}^{-}\right] \approx \Gamma \left[ \mathcal{M%
}\rightarrow D^{0}\overline{D}_{1}^{0}\right] $, and $\Gamma \left[ \mathcal{%
M}\rightarrow D^{-}D_{1}^{+}\right] \approx \Gamma \left[ \mathcal{M}%
\rightarrow \overline{D}^{0}D_{1}^{0}\right] $.

We start from analysis of the channel $\mathcal{M}\rightarrow D^{+}D_{1}^{-}$%
. The three-point correlation function necessary to determine the form
factor $g_{2}(q^{2})$ and hence the strong coupling $g_{2}\equiv \mathcal{Z}%
_{2}(Q^{2}=-m_{D}^{2})$ has the form
\begin{eqnarray}
\Pi _{\mu \alpha \beta }(p,p^{\prime }) &=&i^{2}\int
d^{4}xd^{4}ye^{ip^{\prime }y}e^{-ipx}\langle 0|\mathcal{T}\{\ J_{\mu
}^{D_{1}^{-}}(y)  \notag \\
&&\times J^{D^{+}}(0)J_{\alpha \beta }^{\dagger }(x)\}|0\rangle .
\label{eq:CF1B}
\end{eqnarray}%
In Eq.\ (\ref{eq:CF1B}) $J_{\mu }^{D_{1}^{-}}(x)$ and $J^{D}(x)$ are the
interpolating currents for the mesons $D_{1}^{-}$ and $D^{+}$
\begin{equation}
J_{\mu }^{D_{1}^{-}}(x)=\overline{c}_{i}(x)\gamma _{5}\gamma _{\mu
}d_{i}(x),\ J^{D^{+}}(x)=\overline{d}_{j}(x)i\gamma _{5}c_{j}(x).
\label{eq:Curr1}
\end{equation}

Calculation of $\Pi _{\mu \alpha \beta }^{\mathrm{Phys}}(p,p^{\prime })$ can
be done using the matrix elements
\begin{eqnarray}
\langle 0|J_{\mu }^{D_{1}^{-}}|D_{1}^{-}(p^{\prime },\varepsilon )\rangle
&=&f_{D_{1}}m_{D_{1}}\varepsilon _{\mu },\ \langle 0|J^{D^{+}}|D^{+}\rangle =%
\frac{f_{D}m_{D}^{2}}{m_{c}},  \notag \\
&&
\end{eqnarray}%
and
\begin{equation}
\langle D_{1}^{-}(p^{\prime },\varepsilon )D^{+}(q)|\mathcal{M}(p,\epsilon
)\rangle =g_{2}(q^{2})\epsilon _{\rho \sigma }(p)q^{\rho }\varepsilon
^{\sigma }(p^{\prime }),
\end{equation}%
where $m_{D}$ and $m_{D_{1}}$ are masses of the mesons, whereas $f_{D}$ and $%
f_{D_{1}}$ are their decay constants. Above, by $\varepsilon _{\mu }$ we
denote the polarization vector of axial-vector particle $D_{1}^{-}$.

Then it is not difficult to find that
\begin{eqnarray}
&&\Pi _{\mu \alpha \beta }^{\mathrm{Phys}}(p,p^{\prime })=g_{2}(q^{2})\frac{%
\Lambda f_{D}m_{D}^{2}f_{D_{1}}m_{D_{1}}}{4m_{c}\left( p^{2}-m^{2}\right)
(p^{\prime 2}-m_{D_{1}}^{2})}  \notag \\
&&\times \frac{1}{(q^{2}-m_{D}^{2})}\left\{ g_{\alpha \beta }p_{\mu
}^{\prime }-\frac{(m^{2}+m_{D_{1}}^{2}-q^{2})^{2}}{6m^{2}m_{D_{1}}^{2}}%
g_{\mu \alpha }p_{\beta }^{\prime }\right.  \notag \\
&&\left. +\text{ other contributions}\right\} .  \label{eq:CF8}
\end{eqnarray}

The correlation function $\Pi _{\mu \alpha \beta }^{\mathrm{OPE}%
}(p,p^{\prime })$ is given by the expression
\begin{eqnarray}
&&\Pi _{\mu \alpha \beta }^{\mathrm{OPE}}(p,p^{\prime })=\frac{2}{3}%
g_{\alpha \beta }\int d^{4}xd^{4}ye^{ip^{\prime }y}e^{-ipx}\langle \overline{%
c}c\rangle  \notag \\
&&\times \mathrm{Tr}\left[ \gamma _{\mu }\gamma _{5}S_{d}^{ij}(y)\gamma
_{5}S_{c}^{ja}(-x)S_{c}^{ai}(x-y)\right] .
\end{eqnarray}%
Having used the amplitudes which correspond to structures $g_{\alpha \beta
}p_{\mu }^{\prime }$, we find the SR for the form factor $g_{2}(q^{2})$
\begin{equation}
g_{2}(q^{2})=\frac{4m_{c}(q^{2}-m_{D}^{2})}{\Lambda
f_{D}m_{D}^{2}f_{D_{1}}m_{D_{1}}}%
e^{m^{2}/M_{1}^{2}}e^{m_{D_{1}}^{2}/M_{2}^{2}}\Pi _{2}(\mathbf{M}^{2},%
\mathbf{s}_{0},q^{2}).
\end{equation}%
In computations, the following parameters of the mesons are used \cite%
{PDG:2024,Rosner:2015wva,Gubernari:2022hrq}
\begin{eqnarray}
&&m_{D}=(1869.66\pm 0.05)~\mathrm{MeV},\ f_{D}=(211.9\pm 1.1)~\mathrm{MeV},
\notag \\
&&m_{D_{1}}=(2422.1\pm 0.8)~\mathrm{MeV},\ f_{D_{1}}=180~\mathrm{MeV}.
\end{eqnarray}%
For $M_{2}^{2}$ and $s_{0}^{\prime }$ in the $D_{1}^{-}$ channel we employ
windows
\begin{equation}
M_{2}^{2}\in \lbrack 3.5,4.5]~\mathrm{GeV}^{2},\ s_{0}^{\prime }\in \lbrack
6,8]~\mathrm{GeV}^{2}.
\end{equation}%
The SR data are extracted for $Q^{2}=2-20~\mathrm{GeV}^{2}$ and demonstrated
in Fig.\ \ref{fig:Fit}. The corresponding fit function has parameters $%
\mathcal{Z}_{2}^{0}=2.047$, $z_{2}^{1}=3.591$, and $z_{2}^{2}=-1.888$. For $%
g_{2}$, we find
\begin{equation}
g_{2}\equiv \mathcal{Z}_{2}(-m_{D}^{2})=1.47\pm 0.26.
\end{equation}%
The width of the decay $\mathcal{M}\rightarrow D^{+}D_{1}^{-}$ is calculated
by means of the formula%
\begin{equation}
\Gamma \left[ \mathcal{M}\rightarrow D^{+}D_{1}^{-}\right] =g_{2}^{2}\frac{%
\lambda _{2}}{40\pi m^{2}}|M|^{2},
\end{equation}%
where
\begin{eqnarray}
&&|M|^{2}=\frac{1}{24m^{4}m_{D_{1}}^{2}}\left[
m^{8}-2m^{2}(2m_{D}^{2}-3m_{D_{1}}^{2})\right.  \notag \\
&&\times (m_{D_{1}}^{2}-m_{D}^{2})^{2}+(m_{D_{1}}^{2}-m_{\mathcal{D}%
}^{2})^{4}+m^{6}\left( 6m_{D_{1}}^{2}\right.  \notag \\
&&\left. \left. -4m_{D}^{2}\right)
+2m^{4}(3m_{D}^{4}-8m_{D_{1}}^{2}m_{D}^{2}-7m_{D_{1}}^{4})\right] ,
\end{eqnarray}%
and $\lambda _{2}=\lambda (m,m_{D_{1}},m_{D}).$ For the width of this
process, we get
\begin{equation}
\Gamma \left[ \mathcal{M}\rightarrow D^{+}D_{1}^{-}\right] =(11.9\pm 3.1)~%
\mathrm{MeV}.
\end{equation}

To study the decay $\mathcal{M}\rightarrow D^{-}D_{1}^{+}$ we employ the
correlation function of the following form
\begin{eqnarray}
\widetilde{\Pi }_{\mu \alpha \beta }(p,p^{\prime }) &=&i^{2}\int
d^{4}xd^{4}ye^{ip^{\prime }y}e^{-ipx}\langle 0|\mathcal{T}\{\ J_{\mu
}^{D_{1}^{+}}(y)  \notag \\
&&\times J^{D^{-}}(0)J_{\alpha \beta }^{\dagger }(x)\}|0\rangle ,
\end{eqnarray}%
where $J_{\mu }^{D_{1}^{+}}(x)$ and $J^{D^{-}}(x)$ are relevant
interpolating currents
\begin{equation}
J_{\mu }^{D_{1}^{+}}(x)=\overline{d}_{i}(x)\gamma _{5}\gamma _{\mu
}c_{i}(x),\ J^{D^{-}}(x)=\overline{c}_{j}(x)i\gamma _{5}d_{j}(x).
\label{eq:Curr2}
\end{equation}

Analysis of this decay does not differ from investigations of the decay $%
\mathcal{M}\rightarrow D^{+}D_{1}^{-}$. Therefore, below we write down only
final formulas. Thus the correlator $\widetilde{\Pi }_{\mu \alpha \beta
}(p,p^{\prime })$ in terms of quarks' propagators is given by the expression
\begin{eqnarray}
&&\widetilde{\Pi }_{\mu \alpha \beta }^{\mathrm{OPE}}(p,p^{\prime })=\frac{2%
}{3}g_{\alpha \beta }\int d^{4}xd^{4}ye^{ip^{\prime }y}e^{-ipx}\langle
\overline{c}c\rangle  \notag \\
&&\times \mathrm{Tr}\left[ \gamma _{\mu }\gamma
_{5}S_{c}^{ia}(y-x)S_{c}^{aj}(x)\gamma _{5}S_{d}^{ai}(-y)\right] .
\end{eqnarray}%
The coupling $g_{3}$ is equal to
\begin{equation}
g_{3}\equiv \mathcal{Z}_{3}(-m_{D}^{2})=1.47\pm 0.25.
\end{equation}%
The partial width of the decay $\mathcal{M}\rightarrow D^{-}D_{1}^{+}$
equals to
\begin{equation}
\Gamma \left[ \mathcal{M}\rightarrow D^{-}D_{1}^{+}\right] =(11.7\pm 3.0)~%
\mathrm{MeV}.
\end{equation}%
As is seen, this result coincides with the partial width of the decay $%
\mathcal{M}\rightarrow D^{+}D_{1}^{-}$, because they are connected by $%
c\leftrightarrow \overline{c}$ substitutions. Nevertheless, we have computed
parameters of the mode $\mathcal{M}\rightarrow D^{-}D_{1}^{+}$ explicitly by
confirming the general arguments.


\subsection{Processes $\mathcal{M}\rightarrow D^{+}D^{-}$ and $\ D^{0}%
\overline{D}^{0}$}


The decay mode $\mathcal{M}\rightarrow D^{+}D^{-}$ can be explored by means
of the correlation function
\begin{eqnarray}
\Pi _{\mu \nu }(p,p^{\prime }) &=&i^{2}\int d^{4}xd^{4}ye^{ip^{\prime
}y}e^{-ipx}\langle 0|\mathcal{T}\{J^{D^{+}}(y)  \notag \\
&&\times J^{D^{-}}(0)J_{\mu \nu }^{\dagger }(x)\}|0\rangle ,
\end{eqnarray}%
where the currents $J^{D^{+}}(x)$ and $J^{D^{-}}(x)$ are given by the
expressions Eq.\ (\ref{eq:Curr1}) and
\begin{equation}
\text{ }J^{D^{-}}(x)=\overline{c}_{j}(x)i\gamma _{5}d_{j}(x).
\end{equation}%
The SR for the form factor $g_{4}(q^{2})$ that describes the strong
interaction of particles at the vertex $\mathcal{M}D^{+}D^{-}$ is derived by
computing the correlators $\Pi _{\mu \nu }^{\mathrm{Phys}}(p,p^{\prime })$
and $\Pi _{\mu \nu }^{\mathrm{OPE}}(p,p^{\prime })$ and equating them to get
SR identity.

We find $\Pi _{\mu \nu }^{\mathrm{Phys}}(p,p^{\prime })$ by invoking the
matrix element for the vertex $\mathcal{M}D^{+}D^{-}$
\begin{equation}
\langle D^{+}(p^{\prime })D^{-}(q)|\mathcal{M}(p,\epsilon )\rangle
=g_{4}(q^{2})\epsilon _{\alpha \beta }(p)p^{\prime \alpha }p^{\prime \beta }.
\end{equation}%
After some manipulations, one gets
\begin{eqnarray}
&&\Pi _{\mu \nu }^{\mathrm{Phys}}(p,p^{\prime })=\frac{g_{4}(q^{2})\Lambda
f_{D}^{2}m_{D}^{4}}{m_{c}^{2}\left( p^{2}-m^{2}\right) \left( p^{\prime
2}-m_{D}^{2}\right) \left( q^{2}-m_{D}^{2}\right) }  \notag \\
&&\times \left[ \frac{m^{4}-2m^{2}(m_{D}^{2}+q^{2})+(m_{D}^{2}-q^{2})^{2}}{%
12m^{2}}g_{\mu \nu }\right.  \notag \\
&&\left. +p_{\mu }^{\prime }p_{\nu }^{\prime }+\text{other terms}\right] .
\end{eqnarray}%
For $\Pi _{\mu \nu }^{\mathrm{OPE}}(p,p^{\prime })$, we find%
\begin{eqnarray}
&&\Pi _{\mu \nu }^{\mathrm{OPE}}(p,p^{\prime })=\frac{2}{3}g_{\mu \nu }\int
d^{4}xd^{4}ye^{ip^{\prime }y}e^{-ipx}\langle \overline{c}c\rangle  \notag \\
&&\times \mathrm{Tr}\left[ \gamma _{5}{}S_{d}^{ij}(y)\gamma
_{5}S_{c}^{ja}(-x)S_{c}^{ai}(x-y)\right] .  \label{eq:QCDsideB}
\end{eqnarray}%
To obtain the sum rule for $g_{4}(q^{2})$, we employ the amplitudes $\Pi
_{4}^{\mathrm{Phys}}(p^{2},p^{\prime 2},q^{2})$ and $\Pi _{4}^{\mathrm{OPE}%
}(p^{2},p^{\prime 2},q^{2})$ corresponding to structures $g_{\mu \nu }$ and
obtain
\begin{eqnarray}
&&g_{4}(q^{2})=\frac{12m^{2}m_{c}^{2}(q^{2}-m_{D}^{2})}{\Lambda
f_{D}^{2}m_{D}^{4}[m^{4}-2m^{2}(m_{D}^{2}+q^{2})+(m_{D}^{2}-q^{2})^{2}]}
\notag \\
&&\times e^{m^{2}/M_{1}^{2}}e^{m_{D}^{2}/M_{2}^{2}}\Pi _{4}(\mathbf{M}^{2},%
\mathbf{s}_{0},q^{2}).
\end{eqnarray}

In computations the following parameters have been utilized
\begin{equation}
M_{2}^{2}\in \lbrack 2.5,3.5]~\mathrm{GeV}^{2},\ s_{0}^{\prime }\in \lbrack
4.5,5.5]~\mathrm{GeV}^{2}.
\end{equation}%
The SR data are calculated for $Q^{2}=2-20~\mathrm{GeV}^{2}$ and depicted in
Fig.\ \ref{fig:Fit1}. The extrapolating function $\mathcal{Z}_{4}(Q^{2})$
with $\mathcal{Z}_{4}^{0}=0.949~\mathrm{GeV}^{-1}$, $z_{4}^{1}=1.691$, and $%
z_{4}^{2}=-1.432$ allows one to estimate the coupling $g_{4}$ which reads
\begin{equation}
g_{4}\equiv \mathcal{Z}_{4}(-m_{D}^{2})=(8.09\pm 1.62)\times 10^{-1}\
\mathrm{GeV}^{-1}.
\end{equation}%
The function $\mathcal{Z}_{4}(Q^{2})$ is also plotted in Fig.\ \ref{fig:Fit1}%
.

The partial width of the channel $\mathcal{M}\rightarrow D^{+}D^{-}$ amounts
to
\begin{equation}
\Gamma \left[ \mathcal{M}\rightarrow D^{+}D^{-}\right] =(9.2\pm 2.7)~\mathrm{%
MeV}.
\end{equation}%
The width of second decay$\mathcal{M}\rightarrow D^{0}\overline{D}^{0}$
amounts approximately to $\Gamma \left[ \mathcal{M}\rightarrow D^{+}D^{-}%
\right] $; Related reasons have been previously presented in this section.


\section{ Decays to charmed-strange mesons}

\label{sec:Widths3}

In this section, we are going to consider subdominant channels of the
molecule $\mathcal{M}$ to charmed-strange mesons $D_{s}^{\ast +}D_{s}^{\ast
-}$ , $D_{s}^{+}D_{s1}^{-}(2460)$, and $D_{s}^{-}D_{s1}^{+}(2460)$, and $%
D_{s}^{+}D_{s}^{-}$.


\subsection{ $\mathcal{M}\rightarrow $ $D_{s}^{\ast +}D_{s}^{\ast -}$ and $%
D_{s}^{+}D_{s}^{-}$}


The correlators of the channels $\mathcal{M}\rightarrow $ $D_{s}^{\ast
+}D_{s}^{\ast -}$and $D_{s}^{+}D_{s}^{-}$ can easily be obtained from Eqs.\ (%
\ref{eq:QCDsideA}) and (\ref{eq:QCDsideB}) after replacing $%
S_{d}^{ji}(y)\rightarrow S_{s}^{ji}(y)$. \ For instance, for the decay$%
\mathcal{M}\rightarrow $ $D_{s}^{\ast +}D_{s}^{\ast -}$ we have
\begin{eqnarray}
&&\widetilde{\Pi }_{\mu \nu \alpha \beta }^{\mathrm{OPE}}(p,p^{\prime })=%
\frac{2}{3}g_{\alpha \beta }\int d^{4}xd^{4}ye^{ip^{\prime
}y}e^{-ipx}\langle \overline{c}c\rangle  \notag \\
&&\times \mathrm{Tr}\left[ \gamma _{\mu }S_{s}^{ij}(y)\gamma _{\nu
}S_{c}^{ja}(-x){}S_{c}^{ai}(x-y)\right] .
\end{eqnarray}

It is worth noting that in calculations we take into account terms $\sim
m_{s}=(93.5\pm 0.8)~\mathrm{MeV}$, which appear due to the propagator $%
S_{s}^{ji}(y)$ and the matrix element
\begin{equation}
\langle 0|J^{D_{s}^{\pm }}|D_{s}^{\pm }\rangle =\frac{f_{D_{s}}m_{D_{s}}^{2}%
}{m_{c}+m_{s}}.
\end{equation}%
But, at the same time, we neglect contributions proportional to $m_{s}^{2}$.

The parameters of the mesons $D_{s}^{(\ast )\pm }$ have the following values
\cite{PDG:2024,Rosner:2015wva,Lubicz:2016bbi}
\begin{eqnarray}
m_{D_{s}} &=&(1969.0\pm 1.4)~\mathrm{MeV},\ f_{D_{s}}=(249.0\pm 1.2)~\mathrm{%
MeV},  \notag \\
m_{D_{s}^{\ast }} &=&(2112.2\pm 0.4)~\mathrm{MeV},\ f_{D_{s}^{\ast
}}=(268.8\pm 6.5)~\mathrm{MeV}.  \notag \\
&&
\end{eqnarray}

These decays are characterized by the strong couplings $g_{5}$ and $g_{6}$
at the vertices $\mathcal{M}D_{s}^{\ast +}D_{s}^{\ast -}$ and $\mathcal{M}%
D_{s}^{+}D_{s}^{-}$, respectively. \ In the case of the process $\mathcal{M}%
\rightarrow D_{s}^{\ast +}D_{s}^{\ast -}$ we obtain the following prediction
for $g_{5}$%
\begin{equation}
g_{5}\equiv \mathcal{Z}_{5}(-m_{D_{s}^{\ast }}^{2})=(1.46\pm 0.26)\times
10^{-1}\ \mathrm{GeV}^{-1}.
\end{equation}%
To estimate $g_{5}$, we have employed the extrapolation function $\mathcal{Z}%
_{5}(Q^{2})$ with parameters $\mathcal{Z}_{5}^{0}=0.187~\mathrm{GeV}^{-1}$, $%
z_{5}^{1}=2.037$, and $z_{5}^{2}=-1.467$. The corresponding SR data have
been found using for $M_{2}^{2}$,$\ s_{0}^{\prime }$ in $D_{s}^{\ast }$
channel the regions
\begin{equation}
M_{2}^{2}\in \lbrack 3,5]~\mathrm{GeV}^{2},\ s_{0}^{\prime }\in \lbrack 6,8]~%
\mathrm{GeV}^{2}.
\end{equation}

The coupling $g_{6}$ amounts to
\begin{equation}
g_{6}\equiv \mathcal{Z}_{6}(-m_{D_{s}}^{2})=(5.50\pm 0.96)\times 10^{-1}\
\mathrm{GeV}^{-1},
\end{equation}%
where the function $\mathcal{Z}_{6}(Q^{2})$ is determined by the parameters $%
\mathcal{Z}_{6}^{0}=0.647~\mathrm{GeV}^{-1}$, $z_{6}^{1}=1.525$, and $%
z_{6}^{2}=-1.317$.

The widths of these channels are
\begin{equation}
\Gamma \left[ \mathcal{M}\rightarrow D_{s}^{\ast +}D_{s}^{\ast -}\right]
=(5.8\pm 1.5)~\mathrm{MeV},
\end{equation}%
and
\begin{equation}
\Gamma \left[ \mathcal{M}\rightarrow D_{s}^{+}D_{s}^{-}\right] =(3.6\pm 0.9)~%
\mathrm{MeV},
\end{equation}%
respectively.


\subsection{ $\mathcal{M}\rightarrow $ $D_{s}^{+}D_{s1}^{-}(2460)$ and $%
D_{s}^{-}D_{s1}^{+}(2460)$}


The partial widths of these processes are evaluated using technical methods
and analytical expressions obtained in this article. Because these decays
are connected by $\overline{c}\leftrightarrow c$ replacements they have
similar widths: This fact has been confirmed above explicitly in the case of
other channels. Therefore, we consider only the mode $\mathcal{M}\rightarrow
$ $D_{s}^{+}D_{s1}^{-}$ and employ $\Gamma \left[ \mathcal{M}\rightarrow
D_{s}^{+}D_{s1}^{-}\right] =\Gamma \left[ \mathcal{M}\rightarrow
D_{s}^{-}D_{s1}^{+}\right] $.

The process $\mathcal{M}\rightarrow $ $D_{s}^{+}D_{s1}^{-}$ is characterized
by the strong coupling $g_{7}$ of the particles at the vertex $\mathcal{M}$ $%
D_{s}^{+}D_{s1}^{-}$. Our studies lead to the following result for $g_{7}$
\begin{equation}
g_{7}\equiv \mathcal{Z}_{7}(-m_{D_{s}}^{2})=1.18\pm 0.22.
\end{equation}%
Here, the function $\mathcal{Z}_{7}(Q^{2})$ is determined by the parameters $%
\mathcal{Z}_{7}^{0}=1.692~\mathrm{GeV}^{-1}$, $z_{7}^{1}=3.483$, and $%
z_{7}^{2}=-1.814$. In numerical analysis we have utilized the mass and
decay constant of the meson $D_{s1}^{\pm }(2460)$ \cite%
{PDG:2024,Wang:2015mxa}%
\begin{eqnarray}
m_{D_{s1}} &=&(2459.6\pm 0.9)~\mathrm{MeV},  \notag \\
\ f_{D_{s1}} &=&(345\pm 17)~\mathrm{MeV}.
\end{eqnarray}

The partial width of this decay is
\begin{equation}
\Gamma \left[ \mathcal{M}\rightarrow D_{s}^{+}D_{s1}^{-}\right] =(6.9\pm
1.9)~\mathrm{MeV}.
\end{equation}

Information about partial widths of the dominant and subdominant channels of
the hadronic molecule $\mathcal{M}$ allows us to estimate its full decay
width as

\begin{equation}
\Gamma \lbrack \mathcal{M}]=(149\pm 21)~\mathrm{MeV}.
\end{equation}


\section{Analysis and final remarks}

\label{sec:Conc}

The hadronic tensor molecule $\mathcal{M=}J/\psi J/\psi $ explored in this
article in the context of QCD sum rule method is an unstable structure with
the mass $m=(6290\pm 50)~\mathrm{MeV}$ and width $\Gamma \lbrack \mathcal{M}%
]=(149\pm 21)~\mathrm{MeV}$. The mass of this molecule is comparable with
experimental data for the mass of the resonance $X(6200)$ reported by ATLAS
collaboration
\begin{equation}
6220\pm 50_{-50}^{+40}~\mathrm{MeV.}
\end{equation}

The width of $X(6200)$ measured by the same experimental group amounts to
\begin{equation}
310\pm 120_{-80}^{+70}~\mathrm{MeV,}  \label{eq:ExpW}
\end{equation}%
which is, comparing central values, considerably larger than $\Gamma \lbrack
\mathcal{M]}$. Nevertheless, there is the overlapping region $110-170~%
\mathrm{MeV}$ between the theoretical prediction for $\Gamma \lbrack
\mathcal{M}]$ and experimental information on width of the resonance $X(6200)
$. This fact permits us to interpret the tensor molecule $\mathcal{M}$ as a
candidate to the resonance $X(6200)$. It is possible that the physical state
$X(6200)$ only partly consists of the molecular component $J/\psi J/\psi $,
but this component is evidently important to explain observed features of
the resonance $X(6200)$.

One of the ways to improve the agreement between the theory and experiment
is to include into analysis another decay channels of the molecule $\mathcal{%
M}$. More precise measurements are also necessary to reduce large errors in
Eq.\ (\ref{eq:ExpW}) and make strong conclusions about structure of $X(6200)$%
.

Our studies give also qualitative information on the mass of the excited
molecule $\mathcal{M}(2S)$. In fact, the mass of the ground-state molecule $%
\mathcal{M}$ was obtained using the continuum threshold parameter $s_{0}\in
\lbrack 45,46]~\mathrm{GeV}^{2}$. This means that radially excited molecule
should have a mass $\ m^{2}(2S)>45~\mathrm{GeV}^{2}$ or $m(2S)>6708~\mathrm{%
MeV}$, which excludes the resonance $X(6600)$, but not the next structure $%
X(6900)$. Stated differently, the resonance $X(6900)$ may contain an excited
molecular component. This is only quantitative analysis: For more strong
statements one needs to calculate parameters of $\mathcal{M}(2S)$ which is
beyond the scope of the present article.

The physics of four-charmed tensor mesons is far from being complete. A
credible interpretation of the $X$ resonances is not straightforward and
requires additional detailed investigations.


\begin{thebibliography}{99}

\bibitem{LHCb:2020bwg} R.~Aaij \textit{et al.} (LHCb Collaboration),
Sci.\ Bull. \textbf{65}, 1983 (2020).


\bibitem{ATLAS:2023bft} G.~Aad \textit{et al.} (ATLAS Collaboration),
Phys.\ Rev.\ Lett.\ \textbf{131}, 151902 (2023).


\bibitem{CMS:2023owd} A.~Hayrapetyan \textit{et al.} (CMS Collaboration),
Phys.\ Rev.\ Lett.\ \textbf{132}, 111901 (2024).


\bibitem{CMS:2026tiu} A.~Hayrapetyan \textit{et al.} (CMS Collaboration),
arXiv:2602.02252 [hep-ex].


\bibitem{Anwar:2017toa} M.~N.~Anwar, J.~Ferretti, F.~K.~Guo, E.~Santopinto,
and B.~S.~Zou, Eur.\ Phys.\ J. C \textbf{78}, 647 (2018).


\bibitem{Bedolla:2019zwg} M.~A.~Bedolla, J.~Ferretti, C.~D.~Roberts and
E.~Santopinto,
Eur. Phys. J. C \textbf{80}, 1004 (2020).


\bibitem{Zhang:2020xtb} J.~R.~Zhang,
Phys.\ Rev.\ D \textbf{103}, 014018 (2021).


\bibitem{Wang:2020ols} Z.~G.~Wang,
Chin.\ Phys.\ C \textbf{44}, 113106 (2020).


\bibitem{Albuquerque:2020hio} R.~M.~Albuquerque, S.~Narison,
A.~Rabemananjara, D.~Rabetiarivony, and G.~Randriamanatrika,
Phys.\ Rev.\ D \textbf{102}, 094001 (2020).


\bibitem{Yang:2020wkh} B.~C.~Yang, L.~Tang, and C.~F.~Qiao
Eur.\ Phys.\ J. C \textbf{81}, 324 (2021). 


\bibitem{Cordillo:2020sgc} M.~C.~Gordillo, F.~De~Soto, and J.~Segovia,
Phys.\ Rev.\ D \textbf{102}, 114007 (2020). 


\bibitem{Dong:2020nwy} X.~K.~Dong, V.~Baru, F.~K.~Guo, C.~Hanhart, and
A.~Nefediev,
Phys.\ Rev.\ Lett. \textbf{126}, 132001 (2021); \textbf{127}, 119901(E)
(2021). 


\bibitem{Liang:2021fzr} Z.~R.~Liang, X.~Y.~Wu, and D.~L.~Yao,
Phys.\ Rev.\ D \textbf{104}, 034034 (2021).


\bibitem{Wang:2021kfv} G.~J.~Wang, L.~Meng, M.~Oka, and S.~L.~Zhu,
Phys.\ Rev.\ D \textbf{104}, 036016 (2021). 


\bibitem{Deng:2020iqw} C.~Deng, H.~Chen, and J.~Ping,
Phys.\ Rev.\ D \textbf{103}, 014001 (2021). 


\bibitem{Wang:2022xja} Z.~G.~Wang,
Nucl.\ Phys.\ B \textbf{985}, 115983 (2022). 


\bibitem{Faustov:2022mvs} R.~N.~Faustov, V.~O.~Galkin, and E.~M.~Savchenko,
Symmetry \textbf{14}, 2504 (2022). 


\bibitem{Niu:2022vqp} P.~Niu, Z.~Zhang, Q.~Wang, and M.~L.~Du,
Sci.\ Bull. \textbf{68}, 800 (2023). 


\bibitem{Dong:2022sef} W.~C.~Dong and Z.~G.~Wang,
Phys.\ Rev.\ D \textbf{107}, 074010 (2023). 


\bibitem{Yu:2022lak} G.~L.~Yu, Z.~Y.~Li, Z.~G.~Wang, J.~Lu, and M.~Yan,
Eur.\ Phys.\ J. C \textbf{83}, 416 (2023). 


\bibitem{An:2022qpt} H.~T.~An, S.~Q.~Luo, Z.~W.~Liu, and X.~Liu,
Eur.\ Phys.\ J. C \textbf{83}, 740 (2023). 


\bibitem{Kuang:2023vac} S.~Q.~Kuang, Q.~Zhou, D.~Guo, Q.~H.~Yang, and
L.~Y.~Dai,
Eur.\ Phys.\ J. C \textbf{83}, 383 (2023). 


\bibitem{Liu:2020eha} M.~S.~Liu, F.~X.~Liu, X.~H.~Zhong and Q.~Zhao,
Phys. Rev. D \textbf{109}, 076017 (2024). 


\bibitem{Malekhosseini:2025hyx} M.~Malekhosseini, S.~Rostami, A.~R.~Olamaei
and K.~Azizi,
Nucl. Phys. B \textbf{1018}, 116977 (2025).

\bibitem{Song:2024ykq} Y.~L.~Song, Y.~Zhang, V.~Baru, F.~K.~Guo, and A.~Nefediev, 
Phys.\ Rev.\ D \textbf{111}, 034038 (2025).


\bibitem{Agaev:2023wua} S.~S.~Agaev, K.~Azizi, B.~Barsbay, and H.~Sundu,
Phys.\ Lett.\ B \textbf{844}, 138089 (2023). 


\bibitem{Agaev:2023ruu} S.~S.~Agaev, K.~Azizi, B.~Barsbay and H.~Sundu,
Eur.\ Phys.\ J. Plus \textbf{138}, 935 (2023). 


\bibitem{Agaev:2023gaq} S.~S.~Agaev, K.~Azizi, B.~Barsbay and H.~Sundu,
Nucl.\ Phys.\ A \textbf{1041}, 122768 (2024). 


\bibitem{Agaev:2023rpj} S.~S.~Agaev, K.~Azizi, B.~Barsbay and H.~Sundu,
Eur.\ Phys.\ J. C \textbf{83}, 994 (2023). 


\bibitem{CMS:2025fpt} A.~Hayrapetyan \textit{et al.} (CMS Collaboration),
Nature \textbf{648}, 58 (2025). 


\bibitem{Agaev:2026mif} S.~S.~Agaev, K.~Azizi and H.~Sundu, arXiv:2604.10626 [hep-ph].



\bibitem{Shifman:1978bx} M.~A.~Shifman, A.~I.~Vainshtein and V.~I.~Zakharov,
Nucl.\ Phys.\ B \textbf{147}, 385 (1979).


\bibitem{Shifman:1978by} M.~A.~Shifman, A.~I.~Vainshtein and V.~I.~Zakharov,
Nucl.\ Phys.\ B \textbf{147}, 448 (1979).


\bibitem{Becchi:2020mjz} C.~Becchi, A.~Giachino, L.~Maiani, and
E.~Santopinto,
Phys.\ Lett.\ B \textbf{806}, 135495 (2020).


\bibitem{Becchi:2020uvq} C.~Becchi, A.~Giachino, L.~Maiani, and
E.~Santopinto,
Phys.\ Lett.\ B \textbf{811}, 135952 (2020). 


\bibitem{Agaev:2023ara} S.~S.~Agaev, K.~Azizi, B.~Barsbay, and H.~Sundu,
Phys.\ Rev.\ D \textbf{109}, 014006 (2024). 


\bibitem{Agaev:2020zad} S.~S.~Agaev, K.~Azizi, and H.~Sundu,
Turk.\ J.\ Phys.\ \textbf{44}, 95 (2020). 


\bibitem{PDG:2024} S.~Navas \textit{et al.} [Particle Data Group], Phys.\
Rev.\ D \textbf{110}, 030001 (2024).


\bibitem{Lakhina:2006vg} O.~Lakhina, and E.~S.~Swanson,
Phys.\ Rev.\ D \textbf{74}, 014012 (2006). 


\bibitem{Agaev:2024pil} S.~S.~Agaev, K.~Azizi, and H.~Sundu,
Phys.\ Lett.\ B \textbf{856}, 138886 (2024).


\bibitem{Lucha:2014spa} W.~Lucha, D.~Melikhov, and S.~Simula,
EPJ Web Conf. \textbf{80}, 00043 (2014). 


\bibitem{Rosner:2015wva} J.~L.~Rosner, S.~Stone, and R.~S.~Van de Water,
arXiv:1509.02220.


\bibitem{Gubernari:2022hrq} N.~Gubernari, A.~Khodjamirian, R.~Mandal and
T.~Mannel,
JHEP \textbf{05}, 029 (2022) 


\bibitem{Lubicz:2016bbi} V.~Lubicz, A.~Melis, and S.~Simula,
PoS \textbf{LATTICE2016}, 291 (2017). 


\bibitem{Wang:2015mxa} Z.~G.~Wang,
Eur. Phys. J. C \textbf{75}, 427 (2015) 
\end{thebibliography}
\end{document}